\documentclass[preprint,preprintnumbers,amsmath,amssymb]{revtex4}

\usepackage{graphicx}% Include figure files
\usepackage{dcolumn}% Align table columns on decimal point
\usepackage{bm}% bold math

\begin{document}

\preprint{}
\title{Lattice Gluon Propagator in the Landau Gauge: A Study Using Anisotropic Lattices}
\author{M.~Gong$^a$, Y.~Chen$^b$, G.~Meng$^a$ and C.~Liu$^a$}
\address{
 {\it a}. School of Physics, Peking University, \\
 Beijing 100871, China\\
 {\it b}. Institute of High Energy Physics, Academia Sinica, P.O. Box 918,
 Beijing 100039, China
}

\begin{abstract}
Lattice gluon propagators are studied using tadpole and Symanzik
improved gauge action in Landau gauge. The study is performed using
anisotropic lattices with asymmetric volumes. The Landau gauge
dressing function for the gluon propagator measured on the lattice
is fitted according to a leading power behavior: $Z(q^2)\simeq
(q^2)^{2\kappa}$ with an exponent $\kappa$ at small momenta.  The
gluon propagators are also fitted using other models and the results
are compared. Our result is compatible with a finite gluon
propagator at zero momentum in Landau gauge.
\end{abstract}


\maketitle

\section{Introduce and Motivation}

Quantum Chromodynamics (QCD) is believed to be fundamental theory
for strong interactions in Nature. Almost all information about a
quantum field theory is encoded in the Green's functions and the
knowledge of QCD Green's function is important for the understanding
of some of the novel properties of QCD like confinement and
asymptotic freedom. Gluon propagators have been studied by various
means in the literature. The ultraviolet(UV) behavior of the gluon
propagator can be investigated with perturbation theory. In the
infra-red (IR) region, however, it has to be treated with
non-perturbative methods. One can follow the continuum approaches:
such as truncated Dyson-Schwinger equations(DSEs)~\cite{DSE_intro},
exact renormalization group equations~\cite{RGE_intro}, the
Fokker-Planck type diffusion equation of stochastic
quantization~\cite{SQ_intro}, or the lattice QCD
approaches~\cite{MC_intro1,MC_intro2,MC_intro3,MC_intro4,MC_intro5,MC_intro6,MC_intro7,MC_intro8,ref1,ref2,ref3}.
The results from these two approaches were also compared. Some
agreement were found, however, some important issues remain to be
clarified. In this paper, we study the lattice gluon propagator
using tadpole improved lattice actions on anisotropic lattices.
While most of the previous lattice studies were performed on
isotropic lattices, our results can be compared with both the
previous lattice results and the results using continuum approaches.

In the study of the IR region of the gluon propagator with lattice
QCD, lattice volumes have to be large enough since the minimal
momenta is inverse proportion to the spatial dimensions of the
lattice. Investigations with different lattices imply apparent
finite volume effects~\cite{finite_vol_eff}, and even the extremely
asymmetrical box is not safe~\cite{asym_vol_eff}. A lattice with
large extensions in all four dimensions will require substantial
computational resources. One way to proceed is to adopt a larger
lattice spacing. Previous studies on coarse lattices suggest that
the lattice spacing error is still under control if an improved
lattice action is used~\cite{spacing_eff}. We adopt the tadpole
improved gauge action on anisotropic
lattices~\cite{symanzik1,symanzik2,symanzik3,tadpole}, which has the
advantage of less lattice spacing error and can be used to generate
coarser but larger lattice to reach the deeper IR region. We also
choose to use anisotropic and asymmetrical lattices to further
depress spacing errors and to obtain more low momentum modes.

The anisotropic lattice is a lattice with different cell spacings on
different axes. We set the equal spatial spacing $a_i=a_s$ while the
temporal spacing is $a_t=\frac{1}{\xi_0} a_s$ with the anisotropic
ratio $\xi_0=5$. The anisotropic lattice has the advantage of
further reducing the spacing error while the drawback of further
breaking the four-dimensional Euclidean symmetry. Therefore, the
renormalization effect of anisotropic momenta should be taken into
account which introduces an additional parameter to measure and
complicates the determination of the physical scale.

\section{Lattice Formulations}

\subsection{Gauge Action}

We use the tadpole improved gauge action on anisotropic lattices:
\begin{equation}
S=-\beta\sum_{i>j}(\frac{5}{9}\frac{Tr P_{ij}}{\xi_0 u_s^4} -
\frac{1}{36}\frac{Tr R_{ij}}{\xi_0 u_s^6} - \frac{1}{36}\frac{Tr
R_{ji}}{\xi_0 u_s^6}) - \beta\sum_{i}{(\frac{4}{9}\frac{\xi_0 Tr
P_{0i}}{u_s^2} - \frac{1}{36}\frac{\xi_0 Tr R_{i0}}{u_s^4})}
\label{eq:gauge_action}
\end{equation}
where $P_{ij}$ is the usual plaquette variable and $R_{ij}$ is the
$2\times1$ spatial Wilson loop on the lattice. The parameter $u_s$,
which we take to be the forth root of the average spatial plaquette
value, incorporates the usual tadpole improvement. The parameter
$\xi_0$ designates the bare anisotropy.

\subsection{Gauge Fixing}

The gluon field $\{A_\mu(x)\}$ associated with a gauge configuration
$\{U_\mu(x)\}$ is given by
\begin{equation}
A_\mu(x+a_\mu \hat{e}_\mu / 2) = \frac{1}{2 i
g_0}[U_\mu(x)-U_\mu^\dag(x)] - \frac{1}{6 i g_0} Tr
[U_\mu(x)-U_\mu^\dag(x)] \label{eq:A_mu_x}
\end{equation}

We fix the gluon field to Landau gauge
\begin{equation}
\partial_\mu A_\mu = 0
\label{eq:Landau_gauge}
\end{equation}

To realize the gauge fix, we first define the gauge transform
\begin{equation}
U_\mu^G(x) = G(x)U_\mu(x)G(x+\hat\mu)^\dagger
\label{eq:Landau_gauge_trans}
\end{equation}
where
\begin{equation}
G(x) = e^{-i \sum\limits_\alpha {\omega^\alpha(x)T^\alpha} }
\label{eq:Landau_gauge_transformer}
\end{equation}

Then we define the gauge fixing functionals
\begin{equation}
\begin{aligned}
{\cal F}_1^G[\{U\}]=\sum\limits_{x,\mu}\frac{1}{2}Tr\left\{U_{\mu}^G(x)+U_{\mu}^G(x)^{\dagger}\right\}\\
{\cal F}_2^G[\{U\}]=\sum\limits_{x,\mu}\frac{1}{2}Tr\left\{U_{\mu}^G(x)U_{\mu}^G(x+\hat{\mu})+h.c.\right\}
\label{eq:Landau_gauge_fs}
\end{aligned}
\end{equation}

We adopt the steepest descents method to minimize \cite{cheny_gaugefix}
\begin{equation}
\theta = \frac{1}{2 V} \sum\limits_{x} {Tr (c_1 \Delta_1(x) + c_2 \Delta_2(x))}
\label{eq:Landau_gauge_theta}
\end{equation}

where $c_1$ and $c_2$ are constants tuned to eliminate the artifact in $O(a^2)$ order. The $\Delta_{1,2}$ are defined with the extremum condition of the functionals
\begin{equation}
\begin{aligned}
\frac{\delta{\cal
F}_1^G}{\delta\omega^a(x)}\propto\sum\limits_{\mu}
\left[U_{\mu}^G(x-\hat{\mu})-U_{\mu}^G-
\left(U_{\mu}^G(x-\hat{\mu})-U_{\mu}^G\right)^{\dagger}\right]
\equiv \Delta_1(x)=0 \\
\frac{\delta{\cal F}_2^G}{\delta\omega^a(x)} \propto
\sum\limits_{\mu}Tr\left[U_{\mu}^G(x-2\hat{\mu})U_{\mu}^G(x-\hat{\mu})
-U_{\mu}^G(x)U_{\mu}^G(x+\hat{\mu})-h.c.\right]
\equiv \Delta_2(x)=0
\label{eq:Landau_gauge_delta}
\end{aligned}
\end{equation}

\subsection{The Momentum Space Propagator}

The gluon field in the momentum space can be written as
\begin{equation}
A_\mu(q) = \frac{e^{-i q_\mu / 2}}{2 i
g_0}\{[U_\mu(q)-U_\mu^\dag(-q)] - \frac{1}{3} Tr
[U_\mu(q)-U_\mu^\dag(-q)]\} \label{eq:A_mu_q}
\end{equation}
where $q_\mu$ is the discrete momentum in the periodic
boundary conditions:
\begin{equation}
q_\mu = \frac{2 \pi n_\mu}{a_\mu L_\mu}, \qquad n_\mu = 0, 1
,\ldots,L_\mu - 1 \label{eq:q_mu}
\end{equation}
and $U_\mu(q)$ is the momentum space link, in the form of
\begin{equation}
U_\mu(q) = \sum_{x}{e^{-i q x}
U_\mu(x)}\label{eq:U_mu_q}
\end{equation}

In the continuum, the momentum space propagator in Landau gauge has
the form of
\begin{equation}
D^{ab}_{\mu\nu}(\hat{q}) = \delta^{ab} (\delta_{\mu\nu} - \frac{\hat{q}_\mu
\hat{q}_\nu}{\hat{q}^2}) D(\hat{q}^2) \label{eq:D_cont}
\end{equation}
and therefore the momentum should be corrected as\cite{q_corr}
\begin{equation}
\hat{q}_\mu = \frac{2}{a_\mu}\sin \frac{q_\mu a_\mu}{2}
\label{eq:q_cont}
\end{equation}

The scalar function $D(q^2)$ can be computed on lattice
\begin{subequations}
\begin{eqnarray}
D(\hat{q}^2) = \frac{2}{(N_c^2-1) (N_d-1) V} \sum_{\mu}{\langle
A_\mu(\hat{q}) A_\mu(-\hat{q}) \rangle}, \qquad \hat{q} \neq 0
\label{eq:D_non0}
\\
D(0) = \frac{2}{(N_c^2-1) N_d V} \sum_{\mu}{\langle A_\mu(\hat{q})
A_\mu(-\hat{q}) \rangle}, \qquad \hat{q} = 0 \label{eq:D_0}
\end{eqnarray}
\label{eq:D_all}
\end{subequations}
where $N_c = 3$, $N_d = 4$ are the dimensions of the gauge group and
the space-time, and $V = \prod_{\mu}{L_\mu}$ is the lattice volume.

Another equivalent function is the gluon dressing function:
\begin{equation}
Z(\hat{q}^2) = \hat{q}^2 D(\hat{q}^2) \label{eq:dress_fun}
\end{equation}

\subsection{The Renormalisation of Anisotropic Momenta}

On a symmetric hyper-cubic lattice, the continuum Euclidean symmetry
is broken down to hyper-cubic symmetry. When consider the
low-momentum modes, one normally uses the hyper-cubic lattice
momentum squared defined by:
$\hat{q}^2=(4/a^2)\sum_\mu\sin^2(aq_\mu/2)$ where $a$ being the
lattice spacing. On an anisotropic lattice, however, the hyper-cubic
symmetry is further broken down to cubic symmetry. As a result, when
consider low-momentum modes, one should redefine the anisotropic
momentum squared as:
\begin{equation}
\tilde{q}^2 = \sum \hat{q}_i^2 + (\xi_R \hat{q}_0)^2, \qquad i = 1, 2, 3
\label{eq:renorm_q}
\end{equation}
where the renormalisation coefficient $\xi_R$ is an additional
parameter and can be measured by fitting physical qualities. To tree
level, it is evident that $\xi_R=\xi_0$ with $\xi_0$ being the bare
anisotropy. However, when quantum fluctuations are considered,
$\xi_R$ is in general different from $\xi_0$ and the difference
between the two can be substantial for coarse lattices.

\section{Lattice Simulation}

 Using the pure gauge action~(\ref{eq:gauge_action}), de-correlated
 gauge configurations are generated.
  Three sets of lattices are used in this study and the detailed
simulation parameters are listed in Table \ref{tab:lattices}. For
all lattices, one of the spatial dimensions is twice as large as the
other two which yields more low-momentum modes. The last two sets
have almost equal physical sizes but different lattice spacings so
that the spacing errors and the finite volume effects can be
estimated. The lattice configurations generated are then gauge-fixed
to Landau gauge from which the momentum-space gluon propagators are
measured.

\subsection{Fitting the Infrared Exponent $\kappa$}

 Studies on the dressing function in the Landau gauge using
 Schwinger-Dyson equation (DSE) formalism indicates that the gluon dressing function
 $Z(q^2)$ has a power-law behavior in the IR region:\cite{powerlaw}
\begin{equation}
Z(q^2) \propto (q^2)^{2\kappa} \label{eq:Z_powerlaw}
\end{equation}
The exponent $\kappa$ describes the IR behavior of the gluon
propagator. A $\kappa$ value of $0.5$ indicates that the gluon
propagator is finite, while $\kappa$ values above and below $0.5$
yield infinite and vanishing gluon propagators, respectively.

We choose to fit $Z(\tilde{q}^2)$ measured from our simulations to
estimate the value of $\kappa$, using the MPFIT package which can
perform non-linear regression in IDL environment. The naive errors
are used in these fitting procedures since the correlation between
the gauge configurations is small. Since the power-law model
requires $Z(q^2=0)=0$, data point with zero momentum is discarded.

\begin{figure}
\includegraphics[width=13cm]{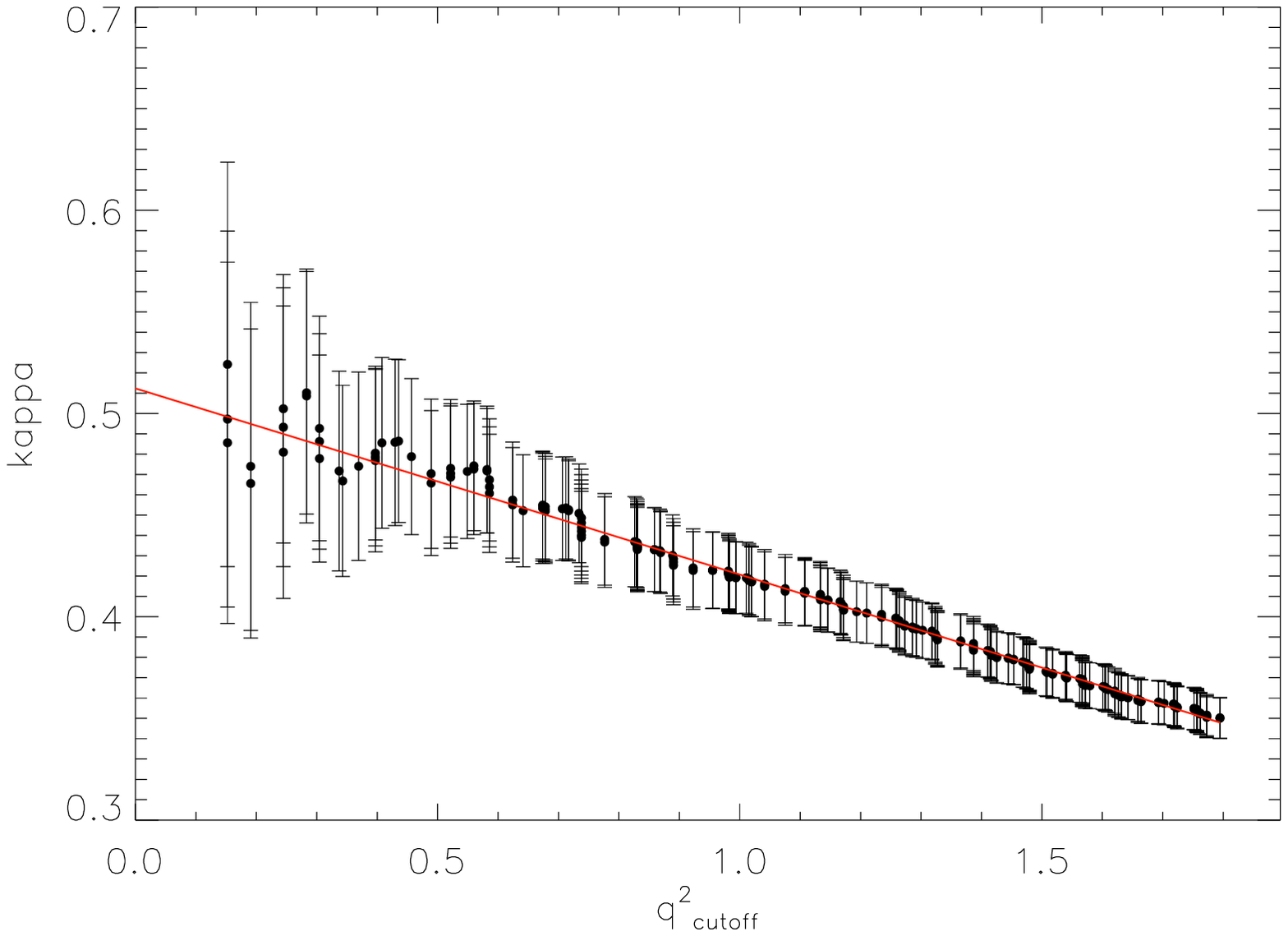}
\caption{\label{fig:renorm_purepowerlaw_16_19} Extrapolation of the
parameter $\kappa$ from a series of fittings with different
$\tilde{q}^2_{\rm cutoff}$ using the power-law model with lattices:
$V = 16 \times 16 \times 32 \times 80$, $\beta = 1.9$.}
\end{figure}

\begin{figure}
\includegraphics[width=13cm]{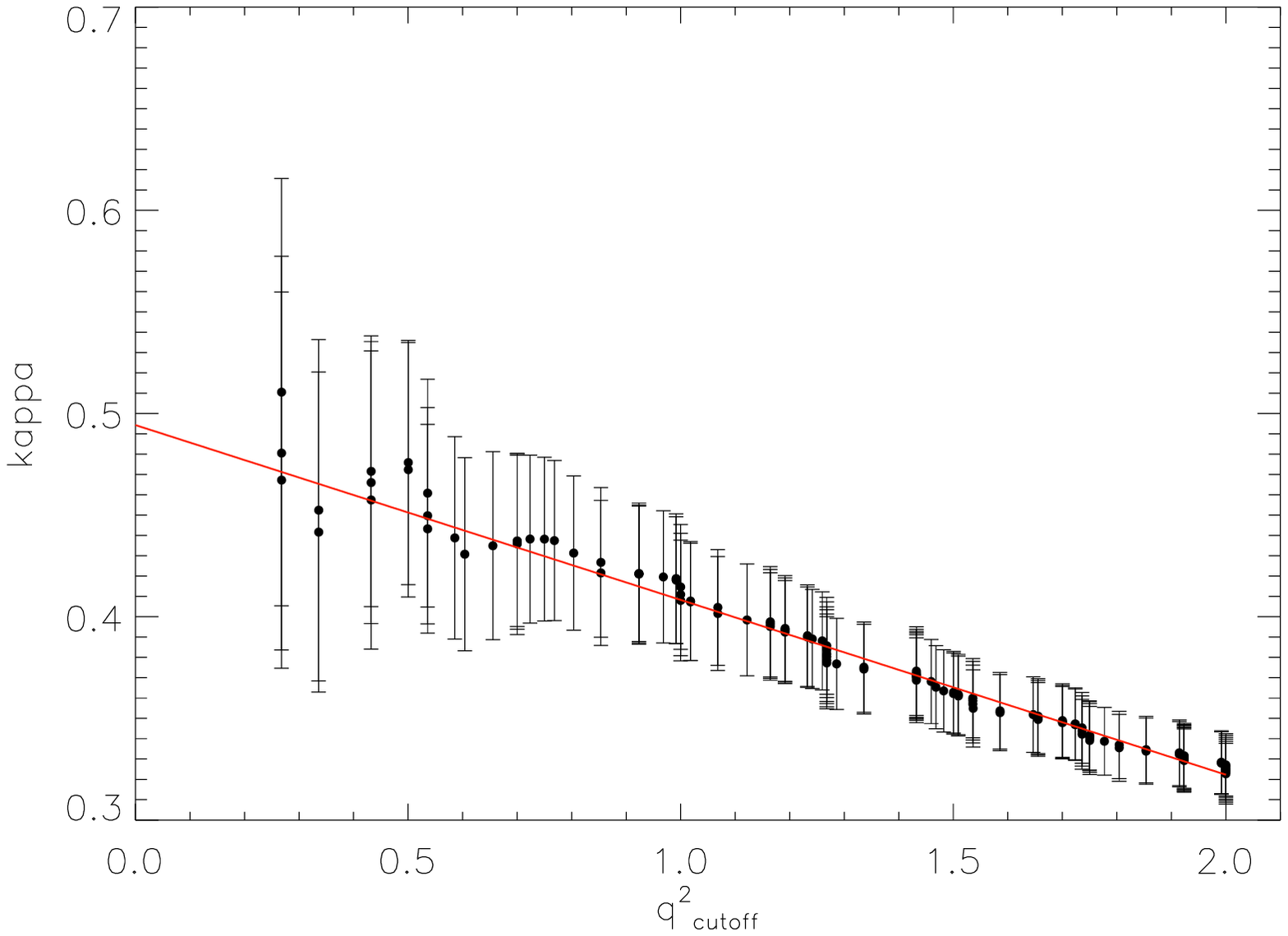}
\caption{\label{fig:renorm_purepowerlaw_12_19} Extrapolation of the
parameter $\kappa$ from a series of fittings with different
$\tilde{q}^2_{\rm cutoff}$ using the power-law model with lattices:
$V = 12 \times 12 \times 24 \times 60$, $\beta = 1.9$.}
\end{figure}

\begin{figure}
\includegraphics[width=13cm]{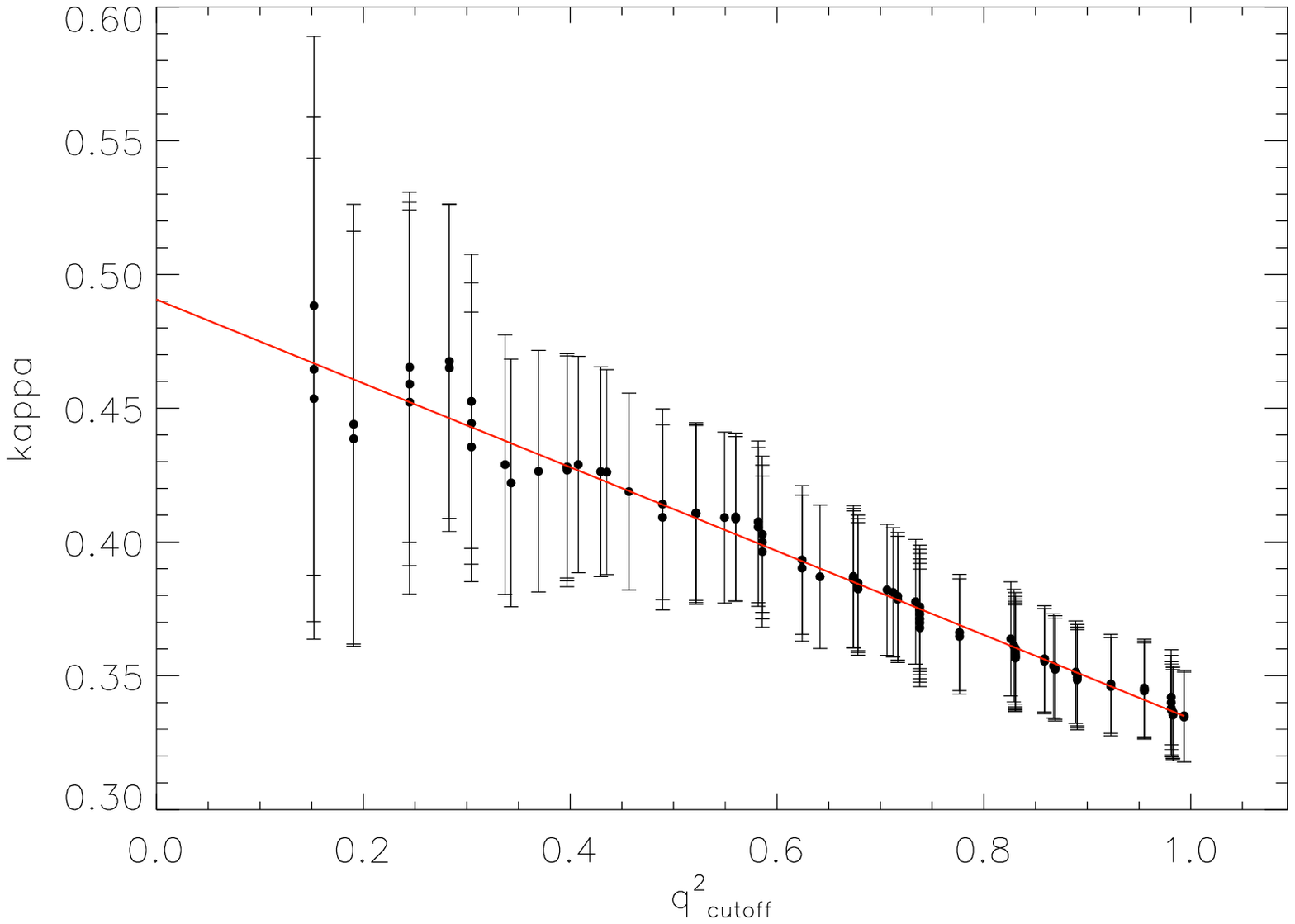}
\caption{\label{fig:renorm_purepowerlaw_16_22} Extrapolation of the
parameter $\kappa$ from a series of fittings with different
$\tilde{q}^2_{\rm cutoff}$ using the power-law model with lattices:
$V = 16 \times 16 \times 32 \times 80$, $\beta = 2.2$.}
\end{figure}

 Since the power-law behavior of $Z(\tilde{q}^2)$ is
 expected to be valid only at small momenta, a cutoff $\tilde{q}^2_{\rm cutoff}>0$
 should be placed on the momentum modes that are measured to set up the
 appropriate fitting range.
 Ideally, if we have enough low-momentum data points, the result of
 the fitting should not depend on the artificial cutoff parameter
 $\tilde{q}^2_{\rm cutoff}>0$.
 But if there were not enough low-momentum data points, the final
 result will have some dependence on the cutoff parameter, which
 turns out to be case of our study. To solve this problem, we
 have tried two methods in the fitting.
 In the first method,
 a series of fittings are performed with variable cutoff parameter $q^2_{\rm cutoff}$.
 The value of $\kappa$ are obtained for each of these fitting ranges.
 The final value of $\kappa$ is determined as a extrapolation in the
 limit $\tilde{q}^2_{\rm cutoff} \rightarrow 0$.
 The results for the value of $\kappa$
 are illustrated in Figure~\ref{fig:renorm_purepowerlaw_16_19},
 \ref{fig:renorm_purepowerlaw_12_19} and
 \ref{fig:renorm_purepowerlaw_16_22} for each
 set of lattice samples. The horizontal axis in these figures
 are the momentum cutoff parameters $\tilde{q}^2_{\rm cutoff}$. The
 data points are the fitted value of $\kappa$ up to the corresponding
 cutoff value. The red curves are the linear extrapolations of
 various values of $\kappa$ obtained at different $\tilde{q}^2_{\rm cutoff}$
 towards the limit $\tilde{q}^2_{\rm cutoff} \rightarrow 0$.
 The extrapolated values at $\tilde{q}^2_{\rm cutoff} \rightarrow 0$
 are the final estimates for $\kappa$ using this fitting method.

\begin{figure}
\includegraphics[width=15cm]{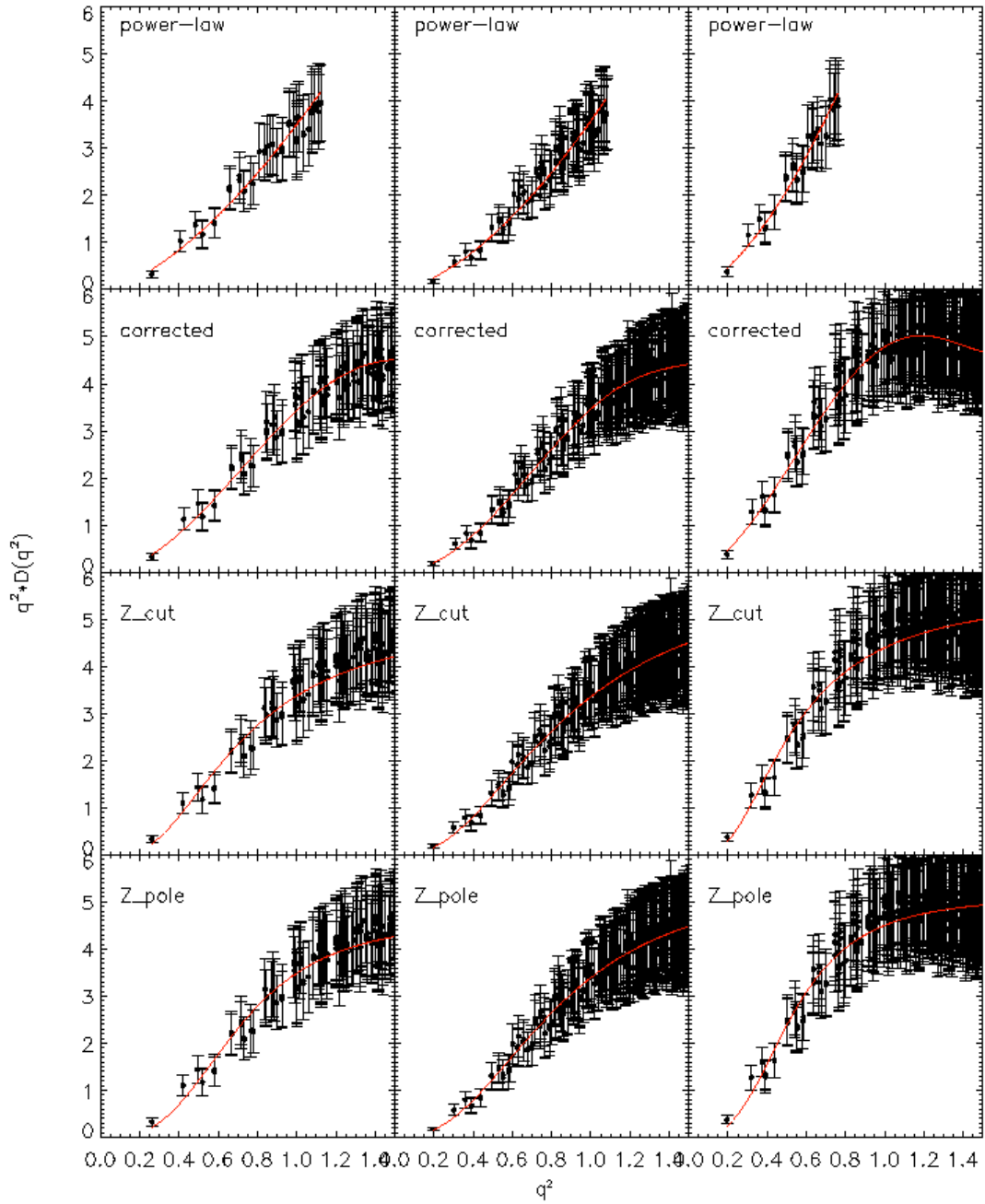}
\caption{\label{fig:fit_all} Fitting the gluon dressing function to
various models.(with Eq.~\ref{eq:Z_powerlaw},
Eq.~\ref{eq:Z_powerlaw_corr}, Eq.~\ref{eq:Z_cut} and
Eq.~\ref{eq:Z_pole} respectively)}
\end{figure}

 In  the first row of the Figure~\ref{fig:fit_all},
 the data points for the dressing function $Z(\tilde{q}^2)$ are plotted together with the
 corresponding power-law behavior fits (the red curve)
 using the extrapolated values of $\kappa$.
 The final extrapolated results for $\kappa$ are also tabulated
 in the first three rows of Table~\ref{tab:fit_results} together with
 the corresponding $\chi^2/d.o.f.$. The values for the
 anisotropy parameter $\xi_R=xi_0$ are obtained using the second
 fitting method to be described below.
 The resulting $\kappa$ values at $\tilde{q}^2_{\rm cutoff} \rightarrow 0$ are consistent
 with a finite propagator at zero momentum, i.e. $\kappa \simeq 0.5$ although
 more definite conclusions will require more accurate simulation results.
 It is also noted that the $\kappa$ values
 obtained from lattices with similar physical sizes (i.e. L12B19 and L16B22)
 but different lattice spacing are consistent with each other within the
 errors, showing that the value of $\kappa$ is not sensitive to
 lattice spacing errors after tadpole improvement.
 The $\kappa$ value from the lattice with a larger physical size
 (i.e. L16B19) is slightly larger from the smaller lattice results by about one standard
 deviation.

 In the fitting method described above,
 the value of $\kappa$ can be estimated quite accurately in the IR region.
 However, since there are not so many low-momentum data points, the value
 of the renormalized anisotropy $\xi_R$ can not be obtained
 accurately. In order to accommodate more data points at larger momenta
 into our fitting procedure, we also adopted a second fitting method.
 In this second fitting method, we express the dressing function
 as the power-law behavior times a polynomial in $\tilde{q}^2$ to
 a certain order:a
\begin{equation}
Z(\tilde{q}^2) \propto (\tilde{q}^2)^{2\kappa} ( 1 + a_1 \tilde{q}^2 +a_2 \tilde{q}^4 + \ldots )
\label{eq:Z_powerlaw_corr}
\end{equation}
 Using this method, the gluon dressing functions for the three set of lattices
 are fitted again. The three plots in the second row of Fig.~\ref{fig:fit_all}
 illustrate the situation of this fitting.
 The results of this fitting procedure are listed in row 4, 5 and 6
 of Table~\ref{tab:fit_results}. The value of $\kappa$ obtained with this method
 is also consistent with a finite propagator at zero momentum, but with a larger
 compared with the results from the previous method.
 However, this method yields more robust values for the anisotropy $\xi_R$. Thus as
 a final step, we fix the $\xi_R$ values obtained from this second method and
 perform the fit of $\kappa$ using the first method. These results
 of $\kappa$ are listed in the first three rows of
 Table~\ref{tab:fit_results}.

\subsection{Fitting Propagator with Other Models}

 There are other ways of parameterizing the dressing function
 $Z(\tilde{q}^2)$~\cite{cut_pole}:
\begin{subequations}
\begin{eqnarray}
Z_{\rm cut}(\tilde{q}^2) \propto (\frac{\tilde{q}^2}{\tilde{q}^2 +
\Lambda^2})^{2 \kappa} \label{eq:Z_cut}
\\
Z_{\rm pole}(\tilde{q}^2) \propto \frac{(\tilde{q}^2)^{2
\kappa}}{(\tilde{q}^2)^{2 \kappa} + (\Lambda^2)^{2 \kappa}}
\label{eq:Z_pole}
\end{eqnarray}
\end{subequations}
 In these parameterizations, another parameter $\Lambda$ is
 introduced.  Obviously, in the small $\tilde{q}^2$ region they
 agree with the previous models. Fitting to these two forms
 are also performed and the corresponding results are given
 in the last six rows of Table~\ref{tab:fit_results}.
 The comparison of the fitted curve together with the data points
 are shown in the last  two rows of Figure~\ref{fig:fit_all}.
 The final results from these models are quite unstable for various lattices
 with large errors to the fitted parameters, both $\kappa$ and $\Lambda$.
 These results seem to be consistent with
 previous studies~\cite{kappa595_1,kappa595_2,kappa595_3}.
 However, since the fitted parameters have large errors and
 the values are not quite consistent on various lattices,
 no definite conclusions should be made from this fitting method.

%\subsection{Fitting the Running Coupling}

%\section{Another Fitting by Using Tree-level Corrected Momenta}
%\subsection{Fitting the Infrared Exponent $\kappa$}
%\subsection{Fitting Propagator with a Cut or a Pole}
%\subsection{Fitting the Running Coupling}

\section{Conclusions}

 In this paper, the gluon propagator in Landau gauge are investigated using
 anisotropic lattices in the IR region. Improved lattice gauge actions are utilized
 which reduces the lattice artifacts substantially. We also used
 lattices with one of the spatial direction is twice as long as the other
 two spatial directions. The largest spatial size we used in this study is about
 $11$fm, allowing us to have more low-momentum modes into the infra-red region.

 The momentum space gluon propagator measured in our lattice
 simulations are fitted using various models.
 The exponent $\kappa$, which characterize the power-law behavior of
 the gluon dressing function in the IR region, is found to be consistent with $0.5$.
 This implies that the gluon propagator may have a finite value at zero
 momentum, in agreement with the result using other non-lattice
 methods. This work also confirms that IR region gluon propagator can be well
 investigated by adopting improved gauge action and lattices with
 large volume and coarse anisotropic spacing.

 \section*{Acknowledgements}
 The authors would like to thank Computer Network Information
 Center (CNIC), Chinese Academy of Sciences and the
 Shanghai Supercomputing Center (SSC) for providing us
 with the computational resources. This work is supported in part by
 the National Science Foundation of China (NSFC) under grant No. 10721063, No. 10675005 and No. 10835002.

\newpage %Just because of unusual number of tables stacked at end

\newpage

\begin{table}
\caption{\label{tab:lattices}Lattice parameters.}
\begin{ruledtabular}
\begin{tabular}{cccccccc}

Lattice&Beta&Physical size&Configurations\\

$16 \times 16 \times 32 \times 80$&1.9&$5.65
\times 5.65 \times 11.30 \times 5.65 fm^4$&200\\
$16 \times 16 \times 32 \times 80$&2.2&$4.47
\times 4.47 \times 8.95 \times 4.47 fm^4$&207\\
$12 \times 12 \times 24 \times 60$&1.9&$4.24
\times 4.24 \times 8.48 \times 4.24 fm^4$&207\\

\end{tabular}
\end{ruledtabular}
\end{table}

\begin{table}
\caption{\label{tab:fit_results}Results for the gluon dressing
function $Z(\tilde{q}^2)$ from various fitting models.}
\begin{ruledtabular}
\begin{tabular}{cccccccc}

Fit Patten                &Parameters  &$\xi_R/\xi_0$        &$\kappa$           &$\Lambda$          &$\chi^2/d.o.f.$\\

Power-law\footnotemark[1] &L16B19      &fixed at $0.6135$   &$0.5124\pm0.0044$  &                   &$0.2054$\\
Power-law\footnotemark[1] &L12B19      &fixed at $0.6541$   &$0.4976\pm0.0097$  &                   &$0.2320$\\
Power-law\footnotemark[1] &L16B22      &fixed at $0.6667$   &$0.4917\pm0.0129$  &                   &$0.2473$\\
Power-law\footnotemark[2] &L16B19      &$0.6135\pm0.0316$   &$0.5117\pm0.0305$  &                   &$0.0725$\\
Power-law\footnotemark[2] &L12B19      &$0.6541\pm0.0261$   &$0.4736\pm0.0292$  &                   &$0.0834$\\
Power-law\footnotemark[2] &L16B22      &$0.6667\pm0.0240$   &$0.4461\pm0.0258$  &                   &$0.0644$\\
$Z_{cut}$                &L16B19      &$0.5832\pm0.0362$   &$0.6745\pm0.0808$  &$0.5493\pm0.1294$  &$0.0782$\\
$Z_{cut}$                &L12B19      &$0.6379\pm0.0306$   &$1.0846\pm0.3258$  &$0.2127\pm0.0868$  &$0.1471$\\
$Z_{cut}$                &L16B22      &$0.6564\pm0.0299$   &$1.0808\pm0.3261$  &$0.1164\pm0.0475$  &$0.1278$\\
$Z_{pole}$               &L16B19      &$0.5874\pm0.0354$   &$0.6128\pm0.0372$  &$0.6913\pm0.0743$  &$0.0736$\\
$Z_{pole}$               &L12B19      &$0.6428\pm0.0293$   &$0.7630\pm0.0639$  &$0.4999\pm0.0335$  &$0.1234$\\
$Z_{pole}$               &L16B22      &$0.6583\pm0.0286$   &$0.7612\pm0.0634$  &$0.2763\pm0.0182$  &$0.1020$\\

\end{tabular}
\end{ruledtabular}
\footnotetext[1]{Extrapolated to zero momenta from a series of
fitting.} \footnotetext[2]{Fitted with Eq.\ref{eq:Z_powerlaw_corr}.}
\end{table}

%\begin{figure}
%\includegraphics[width=13cm]{raw_prop.eps}
%\caption{\label{fig:non_renorm_prop} The gluon propagator with all
%on-axes and off-axes data points.(\ref{eq:D_all}).}
%\end{figure}
%\begin{figure}
%\includegraphics[width=13cm]{raw_dress.eps}
%\caption{\label{fig:non_renorm_dress} The gluon dressing function
%with all on-axes and off-axes data points.(\ref{eq:dress_fun}).}
%\end{figure}

\end{document}